\documentclass[runningheads]{llncs}
\usepackage[T1]{fontenc}
\usepackage{graphicx}
\usepackage{amsmath}
\usepackage{cite}
\usepackage{booktabs}
\usepackage{multirow}
\usepackage{dirtytalk}
\usepackage[colorlinks=true]{hyperref}
\hypersetup{
    linkcolor=red,  
    urlcolor=green, 
}
\usepackage{color, colortbl}
\usepackage[table]{xcolor}

\usepackage{siunitx}
\usepackage[symbol]{footmisc}
\usepackage[misc]{ifsym}

\begin{document}
\title{Diffusion-based Generative Image Outpainting for Recovery of FOV-Truncated CT Images}
\titlerunning{CT-Palette}

\author{Michelle Espranita Liman\inst{1,2} \and Daniel Rueckert\inst{1,3} \and \\ Florian J. Fintelmann$^{\dagger,}$\inst{2,4} \and Philip Müller$^{\dagger,}$\inst{1}\textsuperscript{(\Letter)}}

\authorrunning{M. E. Liman et al.}
\institute{Technical University of Munich, Germany \\
\email{\{michelle.liman,daniel.rueckert,philip.j.mueller\}@tum.de}
\and
Massachusetts General Hospital, USA \\
\email{fintelmann@mgh.harvard.edu}
\and 
Imperial College London, UK
\and
Harvard Medical School, USA}

\footnotetext[2]{Shared last authorship}

\maketitle              
\begin{abstract}
Field-of-view (FOV) recovery of truncated chest CT scans is crucial for accurate body composition analysis, which involves quantifying skeletal muscle and subcutaneous adipose tissue (SAT) on CT slices. This, in turn, enables disease prognostication. Here, we present a method for recovering truncated CT slices using generative image outpainting. We train a diffusion model and apply it to truncated CT slices generated by simulating a small FOV. Our model reliably recovers the truncated anatomy and outperforms the previous state-of-the-art despite being trained on 87\% less data. Our code is available at  \href{https://github.com/michelleespranita/ct\_palette}{https://github.com/michelleespranita/ct\_palette}.

\end{abstract}
\section{Introduction}
Body composition analysis is the analysis of the quantity, quality, and distribution of skeletal muscle and adipose tissue in the body \cite{Magudia2021-ad}. It adds prognostic value to routine CT scans, including for patients with lung cancer \cite{Troschel2020-bn} and pancreatic cancer \cite{Babic2023}. However, chest CT scans are frequently subject to field-of-view (FOV) truncation \cite{Troschel2020-bn}, meaning that part of the chest wall muscle and adipose tissue is intentionally excluded to increase the image quality of the lung tissue \cite{kazerooni2014lung}. Consequently, body composition analysis cannot be accurately conducted on truncated scans, and valuable information is lost.


Recovering FOV-truncated portions of chest CT scans presents a notable challenge with two key issues. Firstly, while recovering the truncated information from the raw data prior to image reconstruction would preserve data fidelity \cite{fournié2019ct, huang2019field, Li2019-ph, Ketola2021-vm}, the raw data are periodically purged from CT scanners, making this approach impractical in clinical practice. Therefore, recovering the truncated part of the anatomy following image reconstruction is more realistic, as performed by the current state-of-the-art method known as S-EFOV \cite{xu2023body}. Secondly, the FOV recovery problem is an ill-posed problem, meaning there exists a distribution of possible outputs for a given input. S-EFOV only generates a single recovered slice given a truncated CT slice. This feature does not allow its users to explore other possible recovered slices in the distribution.

To overcome these limitations, we introduce a novel approach for recovering truncated CT slices using generative image outpainting called \emph{CT-Palette}. Inspired by the recent advancements of generative AI models, CT-Palette is built upon diffusion models \cite{sohldickstein2015deep, ho2020denoising}, which have been proven to outperform generative adversarial networks (GANs) for image-to-image tasks \cite{dhariwal2021diffusion}. Leveraging diffusion models enables CT-Palette to generate multiple recovered slices from a truncated CT slice, making it ideal for tackling the ill-posed FOV recovery problem.

Our contributions are as follows:
\begin{itemize}
    \item We propose \emph{CT-Palette}, a diffusion-based CT field-of-view (FOV) recovery method capable of generating multiple recovered slices from a truncated CT slice. Subsequently, we utilize the muscle and SAT areas from each slice to select the most representative recovered slice as the final output.
    \item We train and evaluate CT-Palette using truncated CT slices generated by simulating real-world FOV truncation.
    \item We design a novel method for mask generation, which reduces the area of the unknown region to be outpainted in the CT slice. This improvement makes training more feasible, particularly on a small dataset.
    \item Despite being trained only on a small dataset, our experiments show that CT-Palette outperforms previous state-of-the-art models.
\end{itemize}

\section{Method}

\begin{figure}[t]
    \centering
        \includegraphics[width=\linewidth]{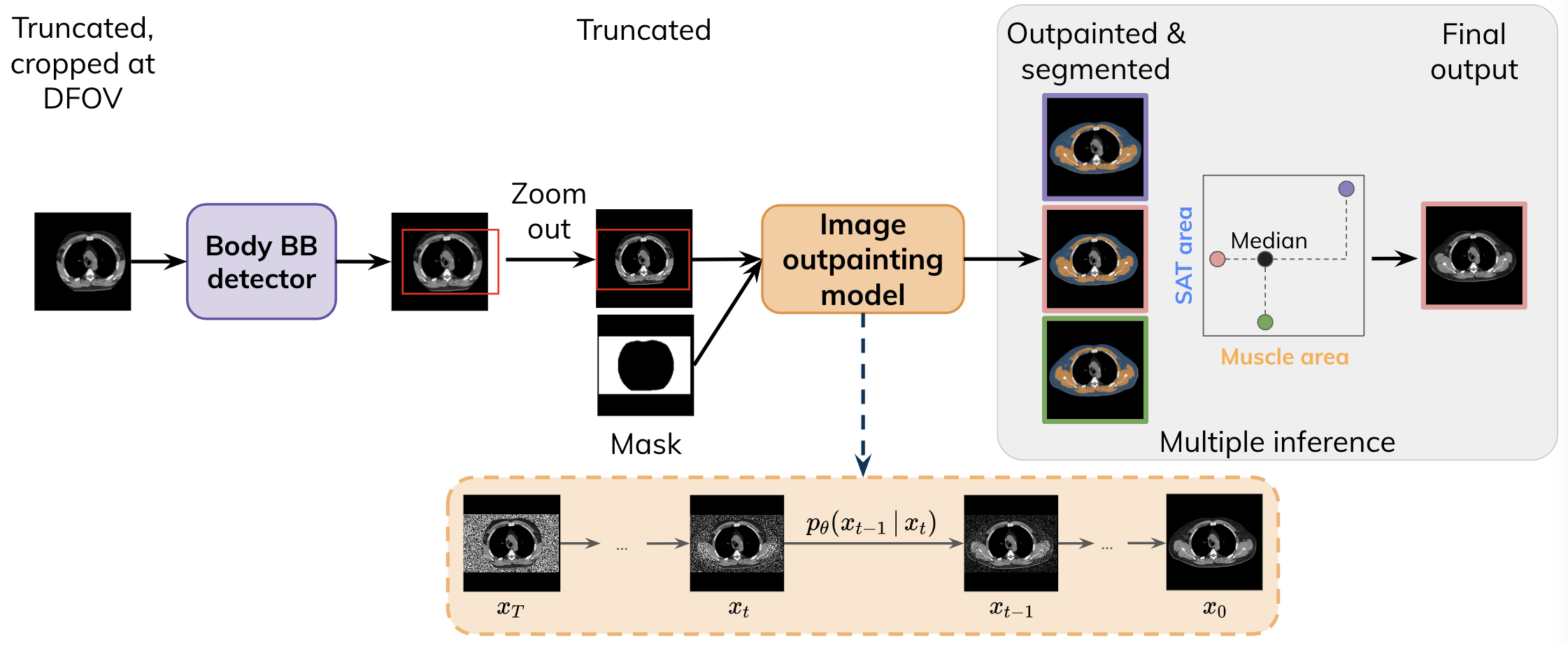}
    \caption{An overview of CT-Palette. The body bounding box detector estimates the bounding box representing the complete/untruncated body, and the image outpainting model recovers the tissues of the truncated CT slice. The white region in the mask indicates the region to be outpainted. CT-Palette generates different slices at each run by drawing samples from the distribution it has learned during training. Using a body composition segmentation model, we extract the muscle and SAT areas from each slice. The final output is the slice with muscle and SAT areas closest to the median.}
    \label{fig:ct_palette_pipeline}
\end{figure}

As shown in Figure \ref{fig:ct_palette_pipeline}, CT-Palette consists of two components: 1) a body bounding box detector which estimates the bounding box representing the untruncated body on the truncated CT slice; 2) an image outpainting model which recovers the tissues of the truncated CT slice. Between the body bounding box detector and the image outpainting model, we \say{zoom out} the slice to ensure the bounding box of the untruncated body fits within the image borders. This ensures sufficient space for the outpainting model to recover the truncated tissues.

In addition to the zoomed-out slice, a binary mask is required as input to the outpainting model. This mask specifies the (unknown) region where the outpainting model should recover the tissues. Typically, this is the FOV mask detected from the truncated CT slice (the pixel values [Hounsfield Unit/HU] located outside the FOV are usually marked by a pre-defined value). However, this approach would cause a large part of the unknown region to include regions outside the body, which is unnecessary for the model to learn.

Therefore, we propose a novel and more effective method to generate masks for the outpainting model, aiming to reduce the size of the unknown region. We refer to the resulting mask as the \say{small mask}. To create the small mask, we use the body bounding box scaled during the zoom-out. Then, we subtract the region occupied by the body (identified using \cite{tang2021bodymask}) from the region covered by the bounding box since the former is a known region that does not need to be learned by the model.

\subsection{Synthetic data generation}

\begin{figure}[ht]
    \centering
        \includegraphics[width=\linewidth]{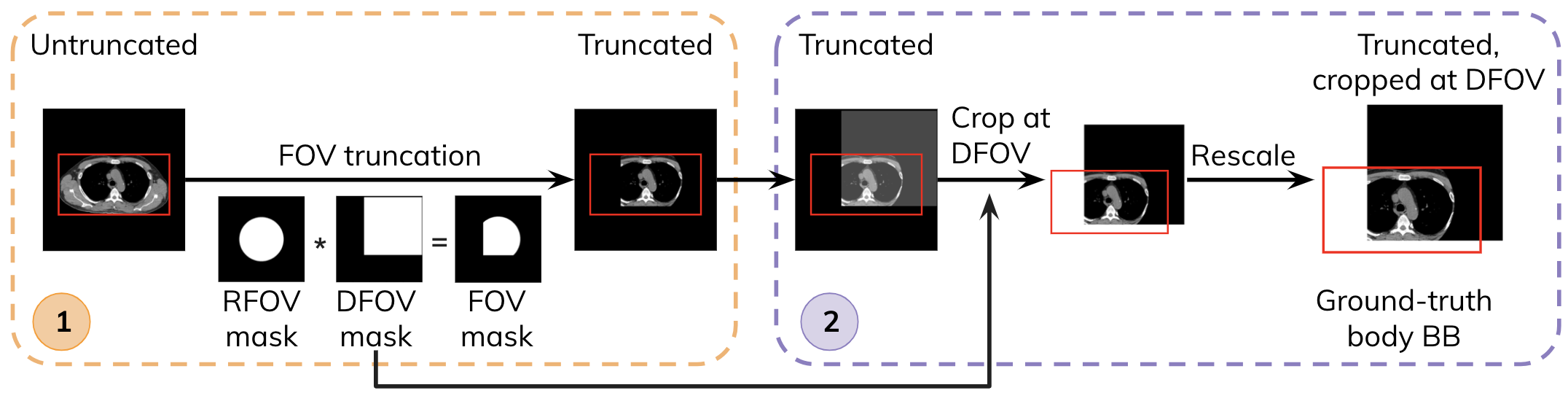}
    \caption{Synthetic data generation process: 1) To train the image outpainting model, we generate truncated slices and their corresponding small masks (explained above); 2) To train the body bounding box detector, we generate truncated slices cropped at the display FOV (DFOV) and bounding boxes of the untruncated bodies.}
    \label{fig:synthetic_data_generation}
\end{figure}

We train the body bounding box detector and the image outpainting model separately (Figure \ref{fig:synthetic_data_generation}). To train the body bounding box detector, we need truncated slices cropped at the display FOV (DFOV) and bounding boxes of the untruncated bodies. To train the image outpainting model, we need truncated slices, their corresponding small masks, and untruncated slices.

To simulate the FOV truncation that occurs in CT scans, we generate synthetic FOV masks by combining the following FOV types \cite{xu2023body}:
\begin{enumerate}
    \item Reconstruction FOV (RFOV): the region of the scanned area used for image reconstruction. We generate a circular mask at the center, defined by the ratio $r_{\text{RFOV}}$ relative to the slice resolution.
    \item Display FOV (DFOV): the region displayed as the final output to the radiologist. We generate a square mask with a length defined by the ratio $r_{\text{DFOV}}$ relative to the slice resolution. It is then offset from the center of the slice by ($x_{\text{DFOV}}$, $y_{\text{DFOV}}$), chosen from the range of [$-\frac{D}{2}$, $\frac{D}{2}$], where $D = d * (1 - r_{\text{DFOV}})$ and $d$ is the slice resolution.
\end{enumerate}

We use the generated FOV masks to truncate the original untruncated slices, resulting in truncated slices for the outpainting model. We extract the bounding boxes of the untruncated bodies by identifying the body using the body mask identification algorithm developed by \cite{tang2021bodymask} and crop the truncated slices at the DFOV for the bounding box detector.

\subsection{Training}

\subsubsection{Body bounding box detector}
This is a ResNet-18 \cite{he2015deep} model pre-trained on ImageNet \cite{imagenet}, with the last fully-connected layer replaced with a fully-connected layer which has 4 output nodes representing the axis-aligned bounding box coordinates of the untruncated body $\hat{B} = $ ($x_{\text{min}}$, $y_{\text{min}}$, $x_{\text{max}}$, $y_{\text{max}}$).

\subsubsection{Image outpainting model}
Inspired by \cite{saharia2022palette}, this is a conditional diffusion model \cite{chen2020wavegrad, saharia2021image} that learns the distribution of possible untruncated slices given a truncated CT slice. At the beginning of the diffusion process, we inject random Gaussian noise exclusively into the unknown region to be outpainted and keep the known region fixed. Over iterations, the model denoises the unknown region, resulting in an outpainted, untruncated slice.

We train for 36 hours on 8 NVIDIA Quadro RTX 8000 GPUs with up to 48 GB of memory assigned to every GPU, resulting in around 340 epochs. We use a total batch size of 128 and optimize the model using the Adam \cite{kingma2017adam} optimizer with a learning rate of \SI{5e-5} and no weight decay. All the other hyperparameters are set to the same values as in \cite{saharia2022palette}.

\subsection{Inference}

As the diffusion process begins with random noise in the unknown region, we obtain different outpainted slices with each run. Based on this, we propose a novel inference method called \emph{multiple inference}: We generate $n>1$ (default: $n=5$) outpainted slices given a truncated slice and select the most representative slice based on statistics. This prevents an outlier from being the final output.

To select the most representative slice, we extract the body composition metrics, i.e.\ muscle and SAT areas, from each outpainted slice using a segmentation model \cite{bridge2022bodycomp} and calculate the median muscle area $\bar{x}_{\text{muscle}}$ and the median SAT area $\bar{x}_{\text{SAT}}$. Then, we calculate the $L_1$ distance from the muscle and SAT areas of each outpainted slice to their corresponding medians:

\begin{equation}
    L_1(x_i, \bar{x}) = |x_{i, \text{muscle}} - \bar{x}_{\text{muscle}}| + |x_{i, \text{SAT}} - \bar{x}_{\text{SAT}}|
\end{equation}

where $x_{i, \{\text{muscle}/\text{SAT}\}}$ is the muscle/SAT area of the outpainted slice $x_i$. Finally, the outpainted slice with the smallest distance to the median is selected.

\section{Experimental Setup}

\subsection{Dataset and Pre-processing}

We use chest CT scans from the Framingham Heart Study \cite{framinghamheartstudy} for training and evaluation. The scans were obtained between 2002 and 2011. After an image quality review by a radiologist, the included patients are 54\% female, 91\% white, with a mean age of 56.1 $\pm$ 10.4 years, and a mean BMI of 29.9 $\pm$ 5.4 kg/m\textsuperscript{2}.

We utilize only slices representing the T5, T8, and T10 vertebral levels of the CT scans, commonly used for body composition analysis on chest CTs \cite{bridge2022bodycomp}. We process the slices by applying a soft-tissue window of range [-160, 240] HU and linear transformation to ensure pixel values range between [-1, 1]. Afterwards, we use the body mask corresponding to the CT slice to remove extraneous objects outside the body, such as the scan table. Finally, the CT slices are downsampled from a resolution of 512 to 256 using bilinear interpolation.

To prevent data leakage, we split the dataset into train (9,061 slices from 3,152 patients), val (899 slices from 177 patients), and test (998 slices from 210 patients) sets by patient ID. There is no significant difference in patient characteristics between the sets.


\subsection{Baselines}
As baselines, we use S-EFOV \cite{xu2023body}, the current state-of-the-art; and RFR-Net \cite{li2020recurrent}, the underlying architecture of S-EFOV's outpainting model. We only compare the outpainting models, as all methods use the same body bounding box detector.

S-EFOV was originally trained on 71,319 slices from the Vanderbilt Lung Screening Program (VLSP) \cite{vlspdataset} dataset, covering vertebral levels slightly above T5 and below T10. To ensure a fair comparison, we use the published weights for S-EFOV with and without fine-tuning on our dataset. For fine-tuning, we adopt the same training settings outlined in \cite{xu2023body}, incorporating data augmentation and utilizing FOV masks as input to the outpainting model.

To highlight the difference in performance between our method and RFR-Net, we train RFR-Net from scratch on our dataset using small masks and the same values for the synthetic data generation parameters ($r_{\text{RFOV}} \sim \mathcal{U}[0.5, 0.7]$ and $r_{\text{DFOV}} \sim \mathcal{U}[0.65, 0.9]$). We optimize RFR-Net using Adam with a learning rate of \SI{2e-4} for a maximum of 200 epochs. Early stopping is implemented with a patience threshold of 60 epochs.

\subsection{Evaluation Strategy} 
\subsubsection{Quantitative Evaluation}
We evaluate the outpainting models based on their ability to recover the truncated slice. This involves ensuring that the muscle and SAT areas of the recovered slice closely align with those of the corresponding ground-truth untruncated slice, measured by RMSE. We extract the muscle and SAT areas using the segmentation model developed by \cite{bridge2022bodycomp} and calculate the segmented areas (cm\textsuperscript{2}) using the pixel spacing obtained from the corresponding DICOM files.

\subsubsection{Qualitative Evaluation}
We qualitatively evaluate the slices recovered by CT-Palette against those by S-EFOV using real-world truncated slices without ground truth. To that end, we randomly select 40 samples for each vertebral level (T5, T8, and T10). Two trained radiologists (A, a radiology resident with 3 years of experience; B, a radiology attending with over 10 years of experience) independently perform the evaluation. Unaware of which recovered slice corresponds to which model, they are tasked with selecting the more realistic recovered slice or choosing \say{no difference} if neither is more realistic than the other.

\section{Results}

\subsection{Quantitative Evaluation}

\begin{figure}[ht]
    \centering
    \includegraphics[width=\textwidth]{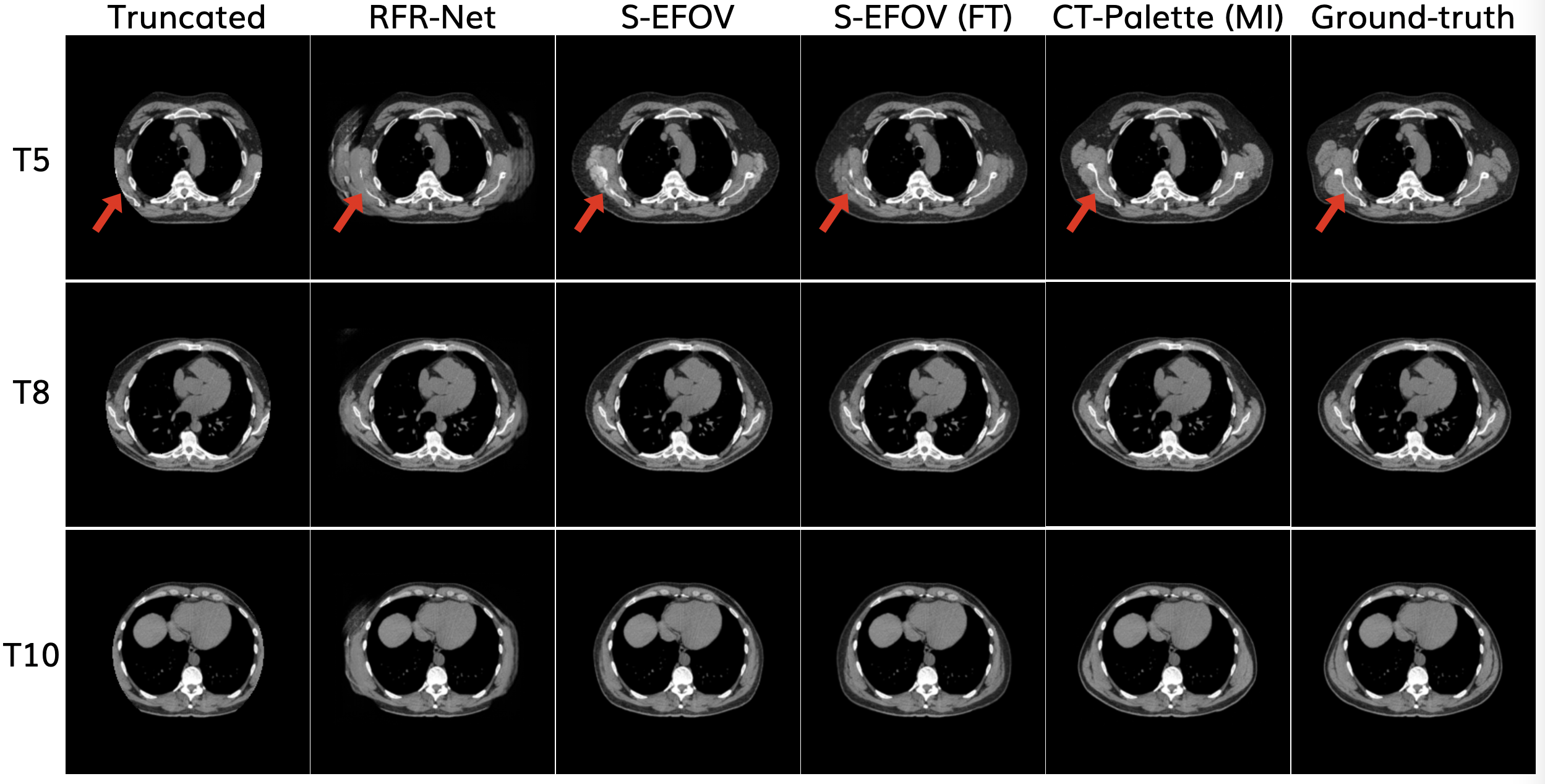}
    \caption{Visual comparison of outpainted slices from different vertebral levels by various models, given a truncated CT slice. (FT): Fine-tuned. (MI): Multiple inference. CT-Palette recovers the truncated slice most realistically. Particularly in the T5 example, CT-Palette successfully restores the shoulder blade, as indicated by the red arrow.}
    \label{fig:visual_comparison}
\end{figure}

Figure \ref{fig:visual_comparison} compares examples of outpainted slices by various models from different vertebral levels. CT-Palette consistently produces realistic slices, while RFR-Net struggles to generate coherent outpaintings despite being trained on the same dataset. Notably, CT-Palette successfully restores the shoulder blade (see red arrow) in the T5 example, which is missed by other models. Additionally, CT-Palette achieves the lowest overall FID \cite{heusel2018gans} (26\% reduction from S-EFOV, see supp. material), indicating that the distribution of its outpainted slices matches that of the ground-truth untruncated slices the most.

\begin{table*}[t!]
    \centering
    \caption{RMSE for muscle and SAT areas between ground-truth untruncated slices and outpainted slices by various models, stratified by vertebral levels. The topmost row indicates the RMSE between untruncated and truncated slices for reference. CT-Palette achieves the biggest reduction in RMSE for both muscle and SAT. (FT): Fine-tuned. (SI): Single inference. (MI): Multiple inference.}
    \label{tab:RMSE_SOTA_v_levels}
    \begin{tabular}{lcccccccc}
        \toprule
        \multirow{2}{*}{Method} & \multicolumn{2}{c}{ Overall (n=997) } & \multicolumn{2}{c}{T5 (n=339)} & \multicolumn{2}{c}{T8 (n=348)} & \multicolumn{2}{c}{ T10 (n=310) } \\
        \cmidrule(lr){2-3}\cmidrule(lr){4-5}\cmidrule(lr){6-7}\cmidrule(lr){8-9} & Muscle & SAT & Muscle & SAT & Muscle & SAT & Muscle & SAT \\
        \midrule 
        Truncated & 26.654 & 76.876 & 38.397 & 83.630 & 21.276 & 80.062 & 12.824 & 64.521 \\
        \midrule
        RFR-Net & 9.022 & 35.258 & 10.778 & 32.388 & 9.134 & 40.776 & 6.412 & 31.376    \\
        S-EFOV & 9.091 &  20.480 & 11.983 & 18.096 & 8.833 & 24.354 & 4.606 & 18.030 \\
        S-EFOV (FT) & 8.202 & 24.974 & 9.014 & 24.252 & 9.188 & 26.653 & 5.723 & 23.773 \\
        CT-Palette (SI) & 5.027 & 16.307 & 5.999 & 16.541 & 5.224 & 17.430 & 3.361 & 14.663 \\
        \rowcolor{lightgray}
        CT-Palette (MI) & \textbf{3.967} & \textbf{12.973} & \textbf{4.381} & \textbf{12.471} & \textbf{4.231} & \textbf{13.467} & \textbf{3.088} & \textbf{12.946} \\
        \bottomrule
    \end{tabular}
\end{table*}

In recovering muscle and SAT areas (Table \ref{tab:RMSE_SOTA_v_levels}), CT-Palette achieves a significant overall reduction in RMSE compared to the best-performing baseline S-EFOV (56.36\% for muscle, 36.66\% for SAT when using multiple inference; 44.70\% for muscle, 20.38\% for SAT when using single inference), despite S-EFOV being pre-trained on a substantially larger dataset.

\begin{figure}[h]
    \centering
        \includegraphics[width=0.9\linewidth]{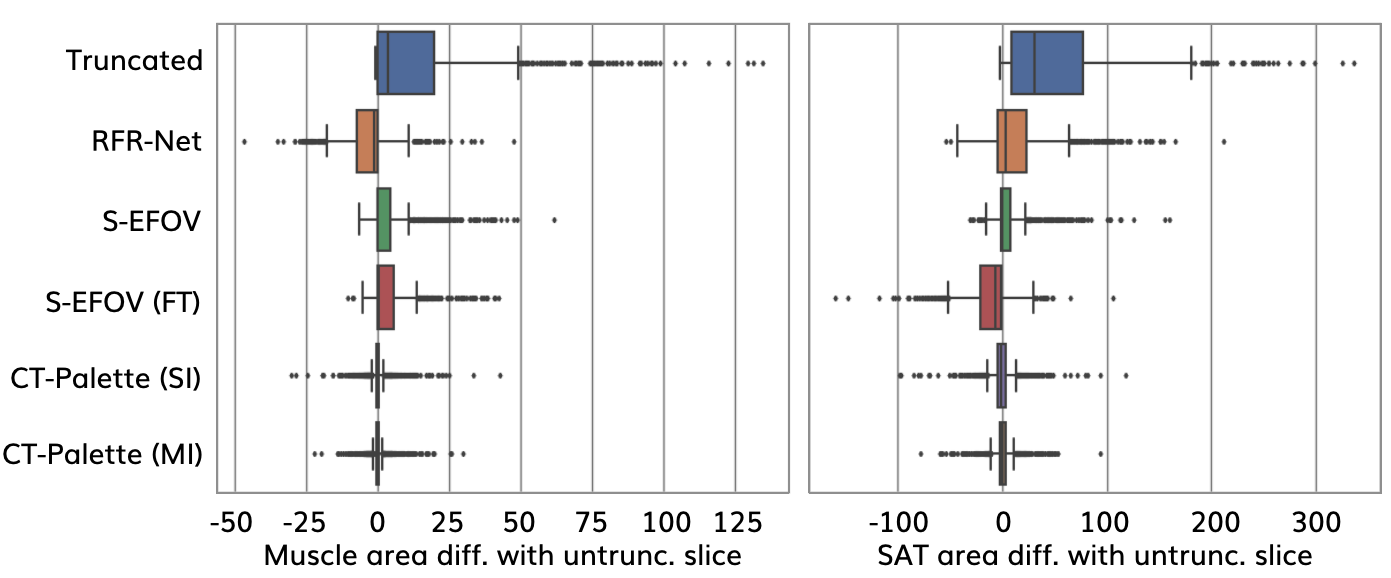}
    \caption{Box plots of the differences in muscle and SAT areas between ground-truth untruncated and outpainted slices by various models. The topmost box plots illustrate the difference in area between the untruncated and truncated slices for reference. Positive values indicate underestimation, negative values indicate overestimation, and 0 indicates no difference. CT-Palette has a median close to zero and very narrow IQR, suggesting little to no deviation from ground-truth in muscle and SAT areas.}
    \label{fig:muscle_sat_boxplot_various_models}
\end{figure}

The box plots in Figure \ref{fig:muscle_sat_boxplot_various_models} illustrate that the differences in muscle and SAT areas between the untruncated and outpainted slices by CT-Palette follow a Gaussian-like distribution, with an equal tendency for over- or underestimation. CT-Palette's box plots exhibit a narrow interquartile range and a median close to zero, indicating no significant difference in area between untruncated and outpainted slices. In contrast, S-EFOV consistently underestimates both muscle and SAT areas, likely due to differences between our dataset and the dataset on which S-EFOV was originally trained. Fine-tuning S-EFOV shifts the distribution of SAT area differences toward a greater inclination for overestimation. The Wilcoxon signed-rank test shows no significant difference in the muscle/SAT area distribution between ground-truth images and those outpainted by our CT-Palette, while other models show a significant difference (see supp. material).

Lastly, our method achieves the highest Dice score overall (0.979 for muscle, 0.954 for SAT when using multiple inference), significantly outperforming S-EFOV by 0.31\% in muscle and 0.63\% in SAT (see supp. material).


We hypothesize that CT-Palette excels even with training only on a small dataset due to its nature as a diffusion model that directly learns the underlying data distribution. In contrast, RFR-Net is designed to minimize a task-specific supervised loss for image inpainting/outpainting. This loss function is composed of style loss, perceptual loss, L1 loss, and TV loss, all of which require extensive experimentation with their coefficients. This inflexibility may hinder RFR-Net's ability to generalize effectively when trained on a limited dataset.

\subsection{Qualitative evaluation}

Radiologist A perceives no significant difference between the CT slices recovered by CT-Palette and S-EFOV in 76.67\% of the samples, while radiologist B finds CT-Palette's slices more realistic than S-EFOV's in 58.33\%. The Cohen's Kappa Coefficient of 0.01 indicates low agreement. Overall, radiologists would tend to find slices from both methods equally realistic, with a slight preference towards those recovered by CT-Palette. The distribution of choices of both radiologists is presented in the supp. material.
 
\section{Discussion and Conclusion}

We presented CT-Palette, a novel approach for recovering truncated chest CT slices via generative image outpainting. It enhances the accuracy of body composition analysis (muscle and SAT areas), thereby enabling disease prognostication. CT-Palette improves on the previous state-of-the-art by generating multiple possible recovered slices given an input. To increase training efficiency on a small dataset, we designed a novel mask generation method that substantially reduces the size of the unknown region for outpainting. CT-Palette demands considerably less data than the previous state-of-the-art while achieving superior performance.

The application of CT-Palette extends beyond chest CT scans to any scan obtained for routine care, including the neck and abdomen. Currently, CT-Palette's main drawback is its slow inference speed due to the large number of steps involved in image generation by the outpainting model. We leave the exploration of techniques to enhance CT-Palette's inference speed for future work.


\begin{credits}
\subsubsection{\ackname}
The Framingham Heart Study is supported by Contract No. HHSN268201500001I from the National Heart, Lung, and Blood Institute (NHLBI) with additional support from other sources. This work was supported by the FHS Core Contract (NHLBI award \#75N92019D00031). This manuscript was not prepared in collaboration with investigators of the Framingham Heart Study and does not necessarily reflect the opinions or conclusions of the Framingham Heart Study or the NHLBI.

\subsubsection{\discintname}
The authors have no competing interests to declare that are relevant to the content of this article.
\end{credits}
%
%
%
\bibliographystyle{splncs04}
\bibliography{Paper-0872}

\end{document}


\title{Supplementary Material for\\ Diffusion-based Generative Image Outpainting for Recovery of FOV-Truncated CT Images}
\titlerunning{CT-Palette}

\author{Michelle Espranita Liman \and Daniel Rueckert \and \\ Florian J. Fintelmann$^{\dagger}$ \and Philip Müller$^{\dagger}$}
\institute{}
\authorrunning{M. E. Liman et al.}



\begin{table*}[h]
    \centering
    \caption{Fréchet Inception Distance (FID) of the distribution of ground-truth untruncated vs. outpainted slices by various models, stratified by vertebral levels. CT-Palette achieves the lowest overall FID.}
    \label{tab:fid}
    \resizebox{0.7\textwidth}{!}{
    \begin{tabular}{lllll}
        \toprule
        Method & Overall (n=997) & T5 (n=339) & T8 (n=348) & T10 (n=310)   \\
        \midrule
        RFR-Net & 17.505 & 25.277 & 21.571 & 19.471   \\
        S-EFOV & 4.809 & 8.106 & 6.416 & 5.405   \\
        S-EFOV (FT) & 3.968 & 6.714 & 5.427 & \textbf{4.921} \\
        CT-Palette (SI) & \textbf{3.521} & \textbf{5.579} & 5.218 & 5.111 \\
        \rowcolor{lightgray}
        CT-Palette (MI) & 3.567 & 5.625 & \textbf{5.189} & 5.191 \\
        \bottomrule
    \end{tabular}}
\end{table*}

\begin{table*}[h]
    \centering
    \caption{The mean and standard deviation of the Dice Similarity Coefficient (DSC) between the muscle and SAT segmentation masks of ground-truth untruncated and outpainted slices by various models. The top row indicates the DSC between untruncated and truncated slices for reference. CT-Palette achieves the highest DSC overall, for both muscle and SAT.}
    \resizebox{\textwidth}{!}{
    \setlength{\tabcolsep}{2pt}
    \begin{tabular}{lcccccccc}
        \toprule
        Method & \multicolumn{2}{c}{ Overall (n=997) } & \multicolumn{2}{c}{ T5 (n=339) } & \multicolumn{2}{c}{ T8 (n=348) } & \multicolumn{2}{c}{ T10 (n=310) } \\
        \cmidrule(lr){2-3}\cmidrule(lr){4-5}\cmidrule(lr){6-7}\cmidrule(lr){8-9} & Muscle & SAT & Muscle & SAT & Muscle & SAT & Muscle & SAT \\
        \midrule
        Truncated & 0.940 $\pm$ 0.083 & 0.865 $\pm$ 0.132 & 0.932 $\pm$ 0.078 & 0.862 $\pm$ 0.114 & 0.933 $\pm$ 0.096 & 0.861 $\pm$ 0.141 & 0.956 $\pm$ 0.070 & 0.874 $\pm$ 0.140 \\
        \midrule
        RFR-Net & 0.957 $\pm$ 0.057 & 0.901 $\pm$ 0.078 & 0.963 $\pm$ 0.037 & 0.906 $\pm$ 0.068 & 0.946 $\pm$ 0.072 & 0.894 $\pm$ 0.086 & 0.962 $\pm$ 0.055 & 0.902 $\pm$ 0.078 \\
        S-EFOV & 0.976 $\pm$ 0.040 & 0.948 $\pm$ 0.060 & 0.977 $\pm$ 0.029 & 0.944 $\pm$ 0.059 & 0.969 $\pm$ 0.052 & 0.944 $\pm$ 0.066 & \textbf{0.982} $\pm$ 0.033 & \textbf{0.958}$\pm$ 0.052 \\
        S-EFOV (FT) & 0.974 $\pm$ 0.043 & 0.947 $\pm$ 0.057 & 0.979 $\pm$ 0.024 & 0.947 $\pm$ 0.051 & 0.965 $\pm$ 0.054	& 0.944 $\pm$ 0.059 & 0.978 $\pm$ 0.044 & 0.950 $\pm$ 0.062 \\
        CT-Palette (SI) & 0.977 $\pm$ 0.038 & 0.950 $\pm$ 0.048 & 0.982 $\pm$ 0.022 & 0.953 $\pm$ 0.045 & 0.971 $\pm$ 0.047 & 0.947 $\pm$ 0.052 & 0.979 $\pm$ 0.040 & 0.952 $\pm$ 0.047 \\
        \rowcolor{lightgray}
        CT-Palette (MI) & \textbf{0.979} $\pm$ 0.036 & \textbf{0.954} $\pm$ 0.045 & \textbf{0.984} $\pm$ 0.019 & \textbf{0.957} $\pm$ 0.038 & \textbf{0.972} $\pm$ 0.045 & \textbf{0.951} $\pm$ 0.050 & 0.980 $\pm$ 0.038 & 0.954 $\pm$ 0.045 \\
        \bottomrule
    \end{tabular}}
    \label{tab:dsc}
\end{table*}

\begin{table*}[!htbp]
    \centering
    \caption{PSNR and SSIM between ground-truth untruncated slices and outpainted slices by various models, stratified by vertebral levels. The top row indicates the PSNR and SSIM between untruncated and truncated slices for reference.}
    \label{tab:PSNR_SSIM}
    \resizebox{\textwidth}{!}{
    \setlength{\tabcolsep}{2pt}
    \begin{tabular}{lcccccccc}
        \toprule
        \multirow{2}{*}{Method} & \multicolumn{4}{c}{ PSNR } & \multicolumn{4}{c}{ SSIM } \\
        \cmidrule(lr){2-5}\cmidrule(lr){6-9} & Overall (n=997) & T5 (n=339) & T8 (n=348) & T10 (n=310) & Overall (n=997) & T5 (n=339) & T8 (n=348) & T10 (n=310) \\
        \midrule 
        Truncated & 34.052 $\pm$ 9.029 & 32.599 $\pm$ 7.631 & 33.729 $\pm$ 8.801 & 36.008 $\pm$ 10.265 & 0.974 $\pm$ 0.025 & 0.972 $\pm$ 0.025 & 0.973 $\pm$ 0.026 & 0.977 $\pm$ 0.025 \\
        \midrule
        RFR-Net & 33.569 $\pm$ 7.339 & 32.119 $\pm$ 7.177 & 33.678 $\pm$ 7.433 & 35.032 $\pm$ 7.101 & 0.969 $\pm$ 0.031 & 0.965 $\pm$ 0.031 & 0.968 $\pm$ 0.032 & 0.974 $\pm$ 0.029 \\
        S-EFOV & \textbf{39.390} $\pm$ 12.657 & \textbf{36.793} $\pm$ 10.389 & \textbf{39.470} $\pm$ 12.820 & \textbf{42.147} $\pm$ 14.066 & \textbf{0.976} $\pm$ 0.026 & \textbf{0.973} $\pm$ 0.026 & \textbf{0.976} $\pm$ 0.027 & \textbf{0.981} $\pm$ 0.026 \\
        S-EFOV (FT) & 38.594 $\pm$ 11.957 & 35.982 $\pm$ 10.695 & 38.700 $\pm$ 11.848 & 41.332 $\pm$ 12.734 & 0.975 $\pm$ 0.027 & 0.972 $\pm$ 0.027 & 0.975 $\pm$ 0.027 & 0.979 $\pm$ 0.026 \\
        CT-Palette (SI) & 36.655 $\pm$ 8.363 & 35.518 $\pm$ 8.170 & 36.497 $\pm$ 8.259 & 38.076 $\pm$ 8.479 & 0.974 $\pm$ 0.028 & 0.972 $\pm$ 0.027 & 0.974 $\pm$ 0.028 & 0.978 $\pm$ 0.027 \\
        \rowcolor{lightgray}
        CT-Palette (MI) & 36.907 $\pm$ 8.467 & 35.702 $\pm$ 8.319 & 36.830 $\pm$ 8.317 & 38.311 $\pm$ 8.583 & 0.975 $\pm$ 0.027 & \textbf{0.973} $\pm$ 0.026 & 0.974 $\pm$ 0.028 & 0.979 $\pm$ 0.026 \\
        \bottomrule
    \end{tabular}}
\end{table*}

\begin{table*}[h]
    \centering
    \caption{p-values from the Wilcoxon signed-rank test comparing the distribution of muscle/SAT areas of ground-truth untruncated and outpainted slices by various models. Higher p-values indicate more similar distributions. p-values below 0.05 are considered significantly different and are marked with *. CT-Palette has the most similar muscle/SAT area distribution with the ground-truth.}
    \label{tab:p_value}
    \resizebox{\textwidth}{!}{
    \setlength{\tabcolsep}{2pt}
    \begin{tabular}{lccccccccc}
        \toprule
        Method & \multicolumn{2}{c}{ Overall (n=997) } & \multicolumn{2}{c}{T5 (n=339)} & \multicolumn{2}{c}{T8 (n=348)} & \multicolumn{2}{c}{ T10 (n=310) } \\
        \cmidrule(lr){2-3}\cmidrule(lr){4-5}\cmidrule(lr){6-7}\cmidrule(lr){8-9} & Muscle & SAT & Muscle & SAT & Muscle & SAT & Muscle & SAT \\
        \midrule
        RFR-Net & \SI{2.0e-63}^* & \SI{8.9e-33}^* & \SI{3.0e-20}^* & \SI{3.7e-12}^* & \SI{5.2e-28}^* & \SI{5.7e-16}^* & \SI{5.5e-18}^* & \SI{1.0e-07}^* \\
        S-EFOV & \SI{4.5e-70}^* & \SI{4.1e-17}^* & \SI{1.7e-34}^* & \SI{1.5e-13}^* & \SI{1.9e-29}^* & \SI{6.3e-07}^* & \SI{3.8e-09}^* & 0.144 \\
        S-EFOV (FT)  & \SI{5.0e-104}^* & \SI{2.3e-95}^* & \SI{2.2e-39}^* & \SI{5.7e-29}^* & \SI{1.6e-38}^* & \SI{1.1e-37}^* & \SI{2.0e-29}^* & \SI{7.8e-34}^*\\
        CT-Palette (SI) & \textbf{0.116} & 0.057 & 0.026 & \textbf{0.406} & 0.198 & 0.876 & 0.042	& 0.005 \\
        \rowcolor{lightgray}
        CT-Palette (MI) & 0.065 & \textbf{0.866} & \textbf{0.303} & 0.348 & \textbf{0.681} & \textbf{0.895} & \textbf{0.050} & \textbf{0.248}\\
        \bottomrule
    \end{tabular}}
\end{table*}

\begin{figure}[ht]
    \centering
    \includegraphics[width=\textwidth]{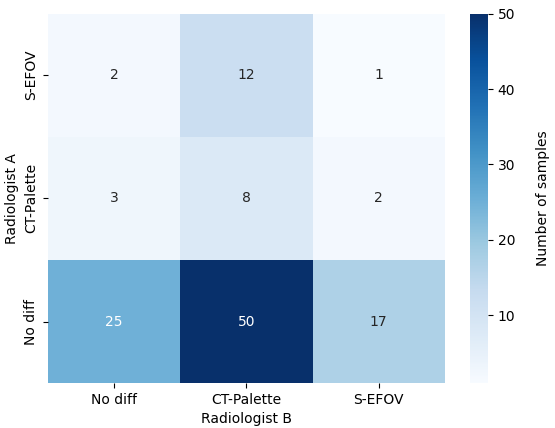}
    \caption{The distribution of choices of two trained radiologists (A and B) on 120 real-world truncated slices (40 slices for each vertebral level T5, T8, and T10) recovered by CT-Palette (MI) and S-EFOV.}
    \label{fig:visual_comparison}
\end{figure}